\documentclass[a4paper]{easychair}

\usepackage{doc}
\usepackage{amsfonts}

\title{Proving UNSAT in SMT:  \\  The Case of Quantifier Free Non-Linear Real Arithmetic}%
\author{
Erika \'Abrah\'am\inst{1}%\thanks{} 
\and
    James H. Davenport\inst{2}\thanks{Support of EPSRC (Grant EP/T015713/1) is gratefully acknowledged.} 
\and
   \\ Matthew England\inst{3}\thanks{Support of EPSRC (Grant EP/T015748/1) is gratefully acknowledged.}
\and
 Gereon Kremer\inst{4}
}

\institute{
RWTH Aachen University, 52062 Aachen, Germany\\
\email{abraham@cs.rwth-aachen.de}\\
\and
   University of Bath,
   Bath BA2 7AY, United Kingdom\\
   \email{J.H.Davenport@bath.ac.uk}\\
\and
   Coventry University, Coventry, CV1 5FB, United Kingdom\\
   \email{Matthew.England@coventry.ac.uk}\\
\and
Stanford University, Stanford, California 94305, United States\\
 \email{gkremer@cs.stanford.edu}
 }

\authorrunning{\'Abrah\'am, Davenport, England and Kremer}

\titlerunning{Proving UNSAT in QF\_NRA}

\begin{document}

\maketitle

\begin{abstract}
We discuss the topic of unsatisfiability proofs in SMT, particularly with reference to quantifier free non-linear real arithmetic.  We outline how the methods here do not admit trivial proofs and how past formalisation attempts are not sufficient.  We note that the new breed of local search based algorithms for this domain may offer an easier path forward.
\end{abstract}

% The table of contents below is added for your convenience. Please do not use
% the table of contents if you are preparing your paper for publication in the
% EPiC Series or Kalpa Publications series

%\setcounter{tocdepth}{2}
%{\small
%\tableofcontents}

%\section{To mention}
%
%Processing in EasyChair - number of pages.
%
%Examples of how EasyChair processes papers. Caveats (replacement of EC
%class, errors).

%------------------------------------------------------------------------------

\section{Introduction}

Since 2013, SAT Competitions have required certificates for unsatisfiability which are verified offline \cite{HJS18}.  As the SAT problems tackled have grown larger, and the solvers have grown more complicated, such proofs have become more important for building trust in these solvers.  The SAT community has agreed on DRAT as a common format for presenting such proofs (although within this there are some flavours \cite{RB19}).  

The SMT community has long recognized the value of proof certificates, but alas producing them turned our to be much more difficult than for the SAT case.  The current version of the SMT-LIB Language (v2.6) \cite{SMTLIB} specifies API commands for requesting and inspecting proofs from solvers but sets no requirements on the form those proofs take.  In fact on page 66 it writes explicitly: ``\emph{The format of the proof is solver-specific}''.  We assume that this is a place holder for future work on an SMT-LIB proof format, rather than a deliberate design.  The paper \cite{BdMF15} summarises some of the requirements, challenges and various approaches taken to proofs in SMT.  Key projects that have been working on this issue include LFSC \cite{Stump2012} and veriT \cite{Barbosa2020}, but there has not been a general agreement in the community yet.

Our long-term vision is that an SMT solver would be able to emit a ``proof'' that covers both the Boolean reasoning and the theory reasoning (possibly from multiple theories) such that a theorem prover (or a combination of multiple theorem provers) could verify its correctness, where the inverted commas indicate that some programming linkage between the theorem provers might be necessary. We would still be some way from having a fully verified one-stop checker as in GRAT \cite{Lammich2020}, but would be a lot closer to it than we are now.

In \cite{BdMF15} the authors explain that since in SMT the propositional and theory reasoning are not strongly mixed, an SMT proof can be an interleaving of SAT proofs and theory reasoning proofs in the shape of a Boolean resolution tree whose leaves are clauses. They identify the main challenge of proof production as keeping enough information to produce proofs, without hurting efficiency too much.  This may very well be true for many cases, but for the area of interest for the authors, \verb+QF_NRA+ (Quantifier-Free Nonlinear Real Arithmetic), there is the additional challenge of providing the proofs of the theory lemmas themselves.

\section{Quantifier Free Non-Linear Real Arithmetic}

\verb+QF_NRA+ typically considers a logical formula $\Phi$ where the literals are statements about the signs of polynomials with rational coefficients, i.e. $f_i(x_1,\ldots,x_n)\sigma_i0$ with $\sigma_i\in\{=,\ne,>,\ge,<,\le\}$.

Any SMT solver which claims to tackle this logic completely relies in some way on the theory of Cylindrical Algebraic Decomposition (CAD).  This was initiated by Collins \cite{Collins1975} in the 1970s with many subsequent developments since:  see for example the collection \cite{CJ98} or the introduction of the recent paper \cite{EBD20}.  The key idea is to decompose infinite space $\mathbb{R}^n$ into a finite number of disjoint regions upon each of which the truth of the constraints is constant.  This may be achieved by decomposing to ensure the signs of the polynomials involved are invariant, although optimisations can produce a coarser, and thus cheaper, decomposition.  

In the case of unsatisfiability an entire CAD truth invariant for the constraints may be produced, and the solver can check that the formula is unsatisfiable for a sample of each cell.  How may this be verified?  The cylindrical condition\footnote{Formally, the condition is that  projection of any two cells onto a lower dimensional space with respect to the variable ordering are either equal or disjoint.  Informally, this means the cells are stacked in cylinders over a decomposition in lower dimensional space.} means that checking our cells decompose the space is trivial, but the fact that the constraints have invariant truth-value is a deep implication of the algorithm, not necessarily apparent from the output.

\subsection*{Past QF\_NRA Formalisation Attempts}

There was a project in Coq to formalise Quantifier Elimination in Real Closed Fields.  This may also be tackled by CAD, and of course has SMT in \verb+QF_NRA+ as a sub-problem.  Work began on an implementation of CAD in Coq with some of the underlying infrastructure formalised \cite{Mahboubi2007}, but the project proceeded to instead formalise QE via alternative methods \cite{CM10}, \cite{CM12b} which are far less efficient\footnote{Although CAD is doubly exponential in the number of variables, the methods verified do not even have worst case complexity bound by a finite tower of exponentials!}.  We learn that the CAD approach was not proven correct in the end \cite[bottom of p. 38]{CM12b}.  Thus while it is formalised that Real QE (and thus satisfiability) is decidable, this does not offer a route to verifying current solver results. 

The only other related work in the literature we found is \cite{NMD15} which essentially formalises something like CAD but only for problems in one variable.

\section{Potential from Coverings Instead of Decompositions?}

There has been recent interaction between the SMT community and the computer algebra community \cite{SC2} from which many of these methods originate.  Computer algebra implementations are being adapted for SMT compliance \cite{SC2}, as CAD was in \cite{KA20}, and there has also be success when they are used directly \cite{FOSKT18}.  Most excitingly, there have been some entirely new algorithmic approaches developed.  
Perhaps most notable is the NLSAT algorithm of Jovanovi\'{c} and de Moura \cite{JdM12}, introduced in 2012 and since generalised into the model constructing satisfiability calculus (mcSAT) framework \cite{dMJ13}.  In mcSAT the search for a Boolean model and a theory model are mutually guided by each other away from unsatisfiable regions. Partial solution candidates for the Boolean structure and for the corresponding theory constraints are constructed incrementally in parallel.  Boolean conflicts are generalised using propositional resolution as normal.  At the theory level, when an assignment (sample point) is determined not to satisfy all constraints then this is generalised from the point to a region containing the point on which the same constraints fail for the same reason.  

In NLSAT, which only considers \verb+QF_NRA+, the samples are generalised to CAD cells\footnote{But not necessarily one that would be produced within any entire CAD for the problem.} being excluded by adding a new clause with the negation of the algebraic description of the cell.  In UNSAT cases these additional clauses become mutually exclusive, in effect the cells generated cover all possible space in $\mathbb{R}^n$.  However, as these are not  arranged cylindrically, this may not be trivial to check from the output.  We note also the more efficient algorithm to compute these single CAD cells in \cite{BK15}, and the new type of decomposition they inspired in \cite{Brown2015}. 

Another new approach was presented recently in \cite{ADEK21}: conflict driven cylindrical algebraic covering (CDCAC).  Like NLSAT this produces a covering of $\mathbb{R}^n$ to show unsatisfiability.  Essentially, a depth first search is performed according to the theory variables. Conflicts over particular assignments are generalised to cells until a covering of a dimension is obtained, and then this covering is generalised to a cell in the dimension below.  In this procedure the covering itself is explicit and easy to verify.  Further, CDCAD computes the covering relative to a set of constraints to check their consistency independent of the Boolean search, meaning it can be more easily integrated into an CDCL(T)-style SMT solver and combined with its other theory solving modules than NLSAT, which is a solving framework on its own.

Both NLSAT and CDCAC rely on CAD theory to conclude that the generalisations of conflicts from models to cells are valid, and so the verification of such theory is still a barrier to verifiable proofs.  But unlike CAD itself, the conflicts that are being generalized are local for both NLSAT and CDCAC.  This may allow a simpler path for verification of individual cases based on the particular relationships of the polynomials involved.  It was observed in \cite{ADEKT20} that a trace of the computation from CDCAC appears far closer to a human derived proof than any of the other algorithms discussed here.  Whether this means it will be more susceptible to machine verification remains to be seen.

\section{Other approaches for QF\_NRA}

We wrote earlier that all solvers tackling \verb+QF_NRA+ in a complete manner rely on CAD based approaches, as these are the only complete methods that have been implemented.  
However, we should note that most solvers also employ a variety of incomplete methods for \verb+QF_NRA+ which tend to be far more efficient than CAD based ones and so are attempted first, and may also be used to solve sub-problems or simplify the input to CAD.  These include incremental linearisation \cite{CGIRS18c}, interval constraint propagation \cite{TVO17}, virtual substitution \cite{Weispfenning1997a}, subtropical satisfiability \cite{FOSV17} and Gr\"obner bases \cite{HEDP16}.  

So, although we think there is potential for verifying output of a cylindrical covering based algorithm, we caution that to obtain fully verified proofs for \verb+QF_NRA+ problems we must take on a greater body of work: to generate proofs for all these methods and furthermore integrate them into or combine them with the CAD proofs.

\label{sect:bib}

\bibliographystyle{alphaurl}
%\bibliographystyle{unsrt}
%\bibliographystyle{abbrv}
%\bibliography{../../../jhd}

\bibliography{mengland}

%------------------------------------------------------------------------------
%------------------------------------------------------------------------------
% Index
%\printindex

%------------------------------------------------------------------------------
\end{document}